\documentclass[aoas,MSNbibl,nameyear,dvips]{arximspdf}
\usepackage{multirow,dcolumn}
\usepackage{graphicx}

\doi{10.1214/12-AOAS605} 
\volume{7}
\issue{2}
\pubyear{2013}
\firstpage{823}
\lastpage{845}

\makeatletter
\newcolumntype{d}[1]{D{.}{.}{#1}}
\newcommand{\eqref}[1]{(\ref{#1})}
\def\mathbbm{\mathbb}
\makeatother

\begin{document}
\begin{frontmatter}

\title{Modeling association between DNA copy number and gene expression
with constrained piecewise linear regression splines}
\runtitle{PLRS for copy number-gene expression association}

\begin{aug}
\author[A]{\fnms{Gwena\"el~G.~R.}~\snm{Leday}\corref{}\thanksref{t1}\ead[label=e1]{gleday@few.vu.nl}},
\author[B]{\fnms{Aad~W.}~\snm{van der Vaart}\ead[label=e2]{avdvaart@math.leidenuniv.nl}},
\author[C]{\fnms{Wessel~N.}~\snm{van~Wieringen}\ead[label=e3]{w.vanwieringen@vumc.nl}}
\and
\author[C]{\fnms{Mark~A.}~\snm{van~de~Wiel}\ead[label=e4]{mark.vdwiel@vumc.nl}}
\runauthor{Leday, van der Vaart, van Wieringen and van de Wiel}
\affiliation{VU University, Leiden University and VU University Medical Center}

\thankstext{t1}{Supported by the Center for Medical Systems Biology (CMSB),
established by the Netherlands Genomics Initiative/Netherlands Organization
for Scientific Research (NGI/NWO).}

\address[A]{G. G.~R. Leday\\
Department of Mathematics\\
VU University\\
De Boelelaan 1081a\\
1081 HV Amsterdam\\
The Netherlands\\
\printead{e1}}

\address[B]{A. W. van der Vaart\\
Mathematical Institute\\
Leiden University\\
PO Box 9512\\
2300 RA Leiden\\
The Netherlands\\
\printead{e2}}

\address[C]{W. N. van Wieringen\\
M. A. van de Wiel\\
Department of Epidemiology and Biostatistics\\
VU University Medical Center\\
PO Box 7057\\
1007 MB Amsterdam\\
The Netherlands\\
\printead{e3}\\
\phantom{E-mail: }\printead*{e4}}

\end{aug}

\received{\smonth{10} \syear{2011}}
\revised{\smonth{8} \syear{2012}}

%
\begin{abstract}
DNA copy number and mRNA expression are widely used data types in
cancer studies, which combined provide more insight than
separately. Whereas in existing literature the form of the
relationship between these two types of markers is fixed
a priori, in this paper we model their association. We employ
piecewise linear regression splines (PLRS), which combine good
interpretation with sufficient flexibility to identify any
plausible type of relationship. The specification of the model leads
to estimation and model selection in a constrained, nonstandard
setting. We provide methodology for testing the effect of DNA on mRNA
and choosing the appropriate model. Furthermore, we present a novel
approach to obtain reliable confidence bands for constrained PLRS, which
incorporates model uncertainty. The procedures are applied to
colorectal and breast cancer data. Common assumptions are found to be
potentially misleading for biologically relevant genes. More flexible
models may bring more insight in the interaction between the two
markers.
\end{abstract}

%

\end{frontmatter}

\section{Introduction}
The genetic material of the human cancer cells often exhibits abnormalities,
of which DNA copy number aberrations are a prime example.
These aberrations comprise gains and losses of chromosome pieces that are
highly variable in size.
Thereby, all or parts of a chromosome may have more or less than the
two copies
received from the parents.
Abnormal DNA copy numbers (different from two) may alter expression
levels of
mRNA transcripts (encoding for functional proteins) that map to the aberration's
genomic location.
Apart from being concordant (copy number tends to correlate positively with
expression level), the form of this association is not established and
may even vary per gene. In this paper we use high-throughput data
available for tissue-specific samples from unrelated patients
to study the relationship between copy number (DNA) and gene expression (mRNA).
We employ a wide class of interpretable models to reflect the
biological mechanism operating between these two molecular levels and
identify relevant markers that may serve as therapeutic
targets.

DNA copy number aberrations are often measured by array comparative genomic
hybridization (aCGH) [\citet{pinkel2005}]. This measuring device is
similar to expression microarrays, which measure expression levels
of thousands of genes simultaneously but interrogate DNA rather than RNA.
Thereby, both profiling experiments produce a continuous value for every
element/probe on the array:
a $\log_{2}$-value of optical fluorescence intensity.
As experiments appear similar, types of information differ and so are
their subsequent treatment. To understand the specific nature of these
data, we include a description of their processing.

\begin{figure}

\includegraphics{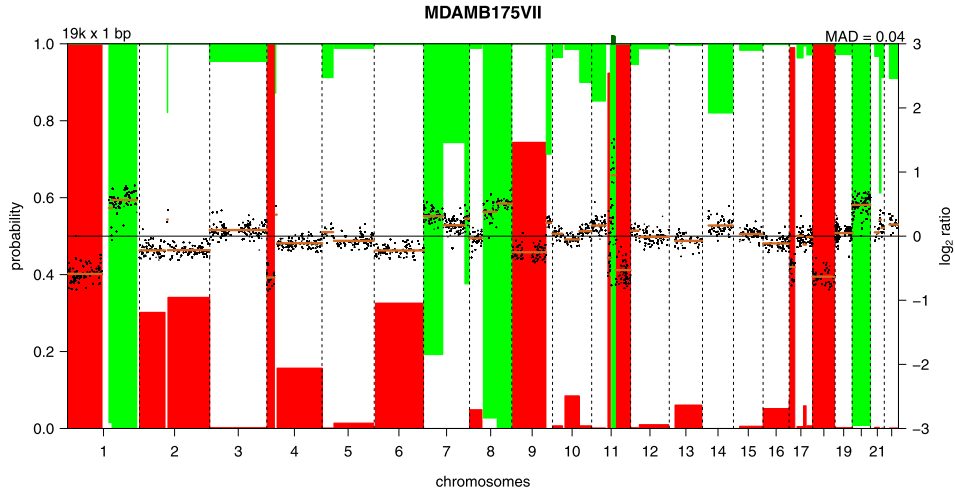}

\caption{Plot of a copy number/aCGH profile from the breast cancer data
set [\citet{neve2006}] showing the different preprocessing steps. Probes
on the array are genomically ordered on the x-axis (only the chromosome
number is displayed). Black dots and orange segments indicate the
normalized and segmented $\log_{2}$-values (right y-axis), respectively.
Bars represent ``loss'' (red) and ``gain'' (green and reversed) membership
probabilities (left y-axis). Amplifications are indicated by tick
marks on
the top axis.} \label{cghcall}
\end{figure}

Normalization of mRNA expression profiles [\citet{quackenbush2002}]
consists in
removing experimental artifacts (such as array differences, means, scales)
and yields, for every gene on each array, a continuous value (normalized
$\log_{2}$-value) which represents the amount of the gene's
transcript present in the sample.
Preprocessing of copy number/aCGH profiles aims to characterize the
genomic instability of each tumor sample and show
deleted/duplicated pieces of chromosomes.
Three successive steps (illustrated in Figure~\ref{cghcall}) are typically
executed to recover the aberration states of all probes [\citet{vandeWiel2010}].
Through these steps, the size, genomic position and type of copy number
aberrations are determined for all samples.
In the first preprocessing step, the \emph{normalization} of $\log_{2}$-values
removes technical or biological artifacts (such as tumor sample contamination,
GC content) and makes the data comparable across samples.
Next, \emph{segmentation} partitions the genome of each sample into
segments of
constant $\log_{2}$-values. These segments are considered a smoothed
(and thus de-noised) version of their normalized counterparts.
Segmentation is motivated by the biological breakpoint process on the
DNA that
may cause differential copy number between neighboring locations.
Finally, \emph{calling} assigns an aberration state to each segment.
Probabilistic calling, usually based on mixture models, results in a
probability distribution over a set of ordered
possible types of genomic aberrations (which we will refer to as states),
typically comprising ``loss'' ($<$2 copies), ``normal'' ($=$2 copies),
``gain'' (3--4 copies) and ``amplification'' ($>$4 copies).
A state is attributed to each probe using a classification rule on the
membership probabilities. Nonprobabilistic calling directly assigns
states to
segmented values, for example, by using a threshold.
Note that larger segmented values almost always correspond to a larger or
equal called copy number (see Figure~\ref{cghcall}).
All in all, the three steps of the preprocessing procedure provide
distinct, but strongly related,
data sets: (1) the normalized, (2)~segmented and (3) called aCGH data.
While most down-stream analyses use either segmented or called data,
we use them jointly.

\begin{figure}

\includegraphics{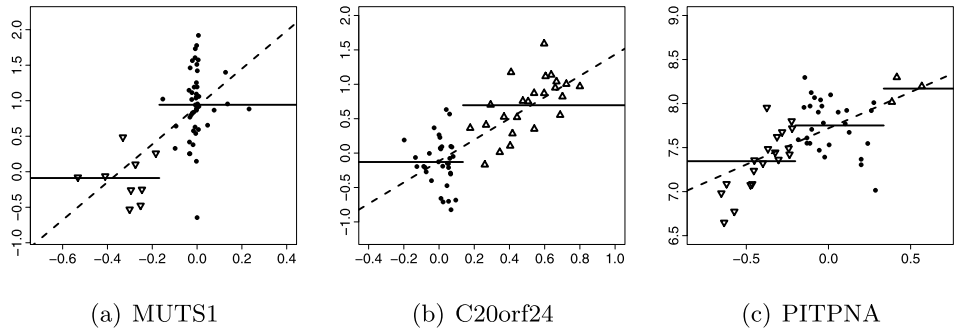}

\caption{Illustration of the association between DNA and mRNA for
three genes in the breast cancer data set [\citet{neve2006}]
used in this study.
Segmented copy number is on the x-axis, while gene expression
is on the y-axis. Symbols indicate the different states, namely,
loss ($\bigtriangledown$), normal ($\bigcirc$) and gain
($\bigtriangleup$). The dashed and ``continuous'' lines
give the fitted linear and stepwise models, respectively.}
\label{ciseffects}
\end{figure}

Current methodology for integrative genomic studies assumes rather
than explores the mathematical form of the relationship between copy number
and expression level. The relationship is said to be either
linear or stepwise (see examples in Figure~\ref{ciseffects}).
A linear relationship is often assumed in combination
with segmented aCGH data. For instance,
the strength of the DNA-mRNA association is measured
by a (modified) correlation coefficient
[\citet{salari2010,schafer2009,lee2008,lipson2004}].
Alternatively, a linear regression
approach is entertained [\citet{asimit2011,menezes2009,gu2008}]. Recently
published multivariate methods [\citet{jornsten2011,peng2010,soneson2010,vanWieringen2010}] also assume linearity. A piecewise DNA-mRNA
relationship is considered when using the called aCGH data for
integrative analysis. \citet{vanWieringen2009} and \citet{bicciato2009} have
proposed stepwise methods.

%

In this paper we develop model selection for piecewise linear regression
splines (PLRS) to decipher how DNA copy number abnormalities
alter the mRNA gene expression level. In addition, we propose a statistical
test that accounts for model uncertainty in the PLRS context to detect those
genes that drive important shifts.
The PLRS framework encompasses the linear and stepwise relationships,
but provides flexibility, while maintaining good interpretability.
In particular, it accommodates \emph{differential} DNA-mRNA
relationships across
states. This is biologically plausible, because the cell has various
post-transcriptional mechanisms to undo the effects of
DNA aberrations. For a given gene, the efficacy of such mechanisms is likely
to differ between gains and losses.
For example, a gain can directly be compensated by regulatory
mechanisms that cause
mRNA degradation, such as methylation. On the other hand, a complete
loss of
both DNA copies (which is more rare than partial loss) cannot be
compensated at all.

Segmented and called data are incorporated into the
analysis, and biologically motivated constraints are imposed on the
model parameters. As this makes model selection and inference
nonstandard, we provide methodology for testing the effect of DNA on
mRNA within the context of PLRS and for selecting the
appropriate model. We also present a novel and computationally
inexpensive method for obtaining uniform confidence bands.
We apply the proposed methodology to colorectal and breast cancer
data sets, where we identify many genes exhibiting nonstandard
behavior.

\section{Methods}
\label{meth}
We model the association between DNA copy number and mRNA expression
by piecewise linear regression splines (PLRS), with biologically
motivated constraints on the coefficients. In this section we address
model selection and describe a modified Akaike criterion in this
context. Further, we present a method for determining uniform
confidence bands, along with a statistical test for the effect of copy
number on mRNA expression.

\subsection{Model}
\label{methmod}
Consider gene expression and aCGH profiling of $n$ independent tumor samples
where for a given gene $\{y_{i},x_{i},s_{i}\}_{i=1}^{n}$ are available,
with $y_{i}$ being the normalized mRNA expression ($\log_{2}$ scale), $x_{i}$
the segmented copy number ($\log_{2}$ scale) and $s_{i}$ the copy
number state (``loss,'' ``normal,'' ``gain'' and ``amplification,'' coded
by $-1$, 0, 1 and 2) value of the $i$th observation, respectively.
Then, the ``full'' model with $S$ states (or parts) takes the form
%
\begin{equation}
\label{eq1} {y_i = f_{\alpha}(x_{i};\theta) +
\varepsilon_{i} = \theta_{0} + \theta_{1}
x_{i} + \sum_{j=1}^{S-1}{\sum
_{d=0}^{1}{\theta_{j,d} (
x_{i}-\alpha_{j} )_{+}^{d}}} +
\varepsilon_{i}}.
\end{equation}
Here
$\theta=\{\theta_{0},\theta_{1},\theta_{1,0},\ldots,\theta_{S-1,0},\theta_{1,1},\ldots,\theta_{S-1,1}\}$
is a vector of $2\times S$ unknown parameters, the $\varepsilon_{i}$ are
independent random variables each normally distributed with
mean $0$ and variance $\sigma^{2}$, and $\{\alpha_{j}\}$ are $S-1$ known
\emph{knots}. The quantity $(a)_{+}^{d}$ represents the positive part
$\max(a,0)$ of $a$ raised to the power $d$. The number of aberration
states $S$
varies across genes. In this study no more than four different aberration
states are considered ($S\leq4$). Below, for the purpose of discussing
model \eqref{eq1} we consider the general case $S=4$.

Knots $\{\alpha_{j}\}$ are obtained using data from the
\emph{calling} preprocessing step. Depending on the type of calling,
two possibilities present themselves.
First, consider nonprobabilistic calling which renders states $\{s_{i}\}_{i=1}^{n}$. Then,
$\alpha_j$ is taken to be the midpoint of the interval between segmented
values $x_i$ belonging to consecutive states (method I). This makes the
(natural) supposition that the calling values respect the ordering of the
segmented values $x_i$ and should be reasonably precise if the
between-state intervals are small, which is typical (see Figure~\ref
{ciseffects}).
Second, consider probabilistic calling, which renders membership (or call)
probabilities: $(p_{i,-1},p_{i,0},p_{i,1},p_{i,2})$. These reflect the
plausibility of the segmented value $x_i$ to belong to the states
$s_i\in\{-1,0,1,2\}$ [\citet{vandeWiel2007}]. Then for $j\in\{1,2,3\}$,
we estimate $\alpha_{j}$ (method II) by
%
\begin{equation}
\label{eq2} {\hat{\alpha}_{j} = \mathop{\operatorname{arg\,max}}_{\alpha\in
\mathbbm{R}} \sum_{i=1}^{n}{p_{i,j(i,\alpha)}}},\qquad
j(i,\alpha) =
\cases{j-2, &\quad $\mbox{if $x_{i} \leq \alpha$},$
\vspace*{2pt}
\cr
j-1, &\quad  $\mbox{if $x_{i} > \alpha$}$.}
\end{equation}
For instance, $\alpha_2$ is the knot between states 0 and 1. To determine
its position, we select for each sample its plausibility
$p_{i,0}$ of belonging to state 0 (when $x_i\le\alpha_2$) or
$p_{i,1}$ of belonging to state 1 (when $x_i>\alpha_2$),
and add over all samples. We select $\alpha_2$ to maximize the sum.
The maximum may not be unique but described by a small
interval; in such a case, we use the corresponding midpoint.
This method may be preferable as it accounts
for the uncertainty of the calling states.
The two methods taken here use data as provided by available
\emph{calling} algorithms.
Proposed models for this preprocessing step typically depend on data
from \emph{all} samples,
which stabilizes the estimation of $\alpha_j$. Furthermore, knots are
to be
interpreted as boundaries between the (ordered) states $\{-1,0,1,2\}$,
which gives us strong a priori knowledge as to their placing (see
Figure~\ref{ciseffects}).
Together, these two arguments support our approach to consider knots
in model \eqref{eq1} as being known. In Section 3 of the supplementary
material (SM) [\citet{leday2012}],
a simulation shows that standard deviations of $\hat{\alpha}_{j}$
are indeed very small.

Model \eqref{eq1} contains seven basis functions besides
the intercept $\theta_0$ and hence is quite flexible. Our approach is
to select appropriate basis functions ($2^{7}=128$ possible models)
and estimate the parameters. The basis functions of degree zero
$x\mapsto(x-\alpha)_+^0$ model discontinuities,
and hence allow for a different effect of copy
number on expression for each state.

This framework is a natural fundament to test meaningful hypotheses. For
example, the hypothesis that for a given state
there is an effect of copy number on mRNA can be expressed
in terms of a linear function of the parameters being zero ($\sum_{j}{\theta_{j,1}}=0$);
a difference between the effects of two adjacent states corresponds to
knot deletion. The submodel consisting of piecewise
constant functions
[without the functions $x\mapsto x$ and $x\mapsto(x-\alpha)_+^1$]
allows testing the difference in expression between
states based on discrete genomic information.

To increase biological plausibility, aid interpretation and increase
the stability of estimation, we impose a set of linear constraints on
the parameters. As it is generally believed that direct causal
effects of DNA on mRNA should be positive, we constrain all slopes
to be nonnegative. More exactly, we constrain the slope corresponding
to the ``normal'' state to be nonnegative ($\theta_{1}+\theta_{1,1}\geq0$),
while others are forced to be at least equal to the latter
(implied by $\theta_{1,1}\leq0$ for losses, $\theta_{2,1}\geq0$ for
gains and $\theta_{2,1}+ \theta_{3,1} \geq0$ for amplifications).
For the same reason we constrain jumps $\theta_{j,0}$
from state to state to be nonnegative.
Note that the restrictions adopted here force the slope of the ``normal''
state to be small or null and make the natural assumption that a
normal copy number is not expected to affect (at least severely)
gene expression.

The maximum likelihood estimator of the unknown vector of coefficients
$\theta$
solves the following convex optimization problem:
%
\begin{equation}
\label{opt1} \mathop{\mathrm{minimize}}_{\theta} (y-\mathbf{X}\theta
)^{T} (y-\mathbf{X}\theta ) \qquad\mbox{subject to } \mathbf{C}\theta\geq0 .
\end{equation}
This can be solved by quadratic programming
[\citet{boyd2004}]. The vector $y=\{y_{1}, \ldots,y_{n}\}$ denotes the
expression signature of a given gene and~$\mathbf{X}$ the
associated matrix of covariates designed according to \eqref{eq1}. The
full row-rank matrix $\mathbf{C}$ expresses the constraints that
are imposed on the parameters. For the 4-state full model we define
$\mathbf{C}$
as the matrix in
%
\begin{equation}
\label{constr} \pmatrix{ 0 & 1 & 0 & 1 & 0 & 0 & 0 & 0 \vspace*{2pt}
\cr
0 & 0 &
0 & 0 & 0 & 1 & 0 & 0 \vspace*{2pt}
\cr
0 & 0 & 0 & 0 & 0 & 1 & 0 & 1
\vspace*{2pt}
\cr
0 & 0 & 0 & -1 & 0 & 0 & 0 & 0 \vspace*{2pt}
\cr
0 & 0 & 1 & 0 & 0
& 0 & 0 & 0 \vspace*{2pt}
\cr
0 & 0 & 0 & 0 & 1 & 0 & 0 & 0 \vspace*{2pt}
\cr
0 & 0
& 0 & 0 & 0 & 0 & 1 & 0 } \pmatrix{ \theta_{0} \vspace*{2pt}
\cr
\theta_{1} \vspace*{2pt}
\cr
\theta_{1,0} \vspace*{2pt}
\cr
\theta_{1,1} \vspace*{2pt}
\cr
\theta_{2,0} \vspace*{2pt}
\cr
\theta_{2,1} \vspace*{2pt}
\cr
\theta_{3,0} \vspace*{2pt}
\cr
\theta_{3,1} } \geq 0.
\end{equation}

\subsection{Model selection}
\label{methmodSel}
Given $R$ competing statistical models, with log-likelihoods
$\mathcal{L}_{r}(\theta_{r})$ based on a $k_{r}\times1$
parameter vector $\theta_{r}$ and with corresponding
maximum likelihood estimators (MLE) $\hat{\theta}_{r}$,
the \emph{Akaike information criterion} (AIC)
selects as best the model that minimizes
%
\begin{equation}
\label{eq3} \mathrm{AIC}_{r}=-\mathcal{L}_{r}(\hat{
\theta}_{r})+k_{r}.
\end{equation}
This information criterion consists of two parts: the negative
maximized log-likelihood, which measures the lack of model fit,
and a penalty for model complexity.
Although AIC has found wide application, it is less suitable for
models that include parameter constraints, as in our situation.
It can be adapted as follows.

The original motivation for the criterion [\citet{akaike1973}] is to choose
the model that minimizes the Kullback--Leibler (KL) divergence to the
true distribution of the data. Indeed, the criterion $\mathrm{AIC}_r$
is (under some conditions) an asymptotically unbiased estimator
of this KL divergence. The likelihood at a given
parameter is an unbiased estimate of the KL divergence at this parameter,
but evaluating it at the maximum likelihood estimator
introduces a bias caused by ``using the data twice,'' which is
compensated by the penalty $k_r$ [\citet{bozdogan1987}].
In the constrained case (i.e., subject to $\mathbf{C}\theta\geq0$)
we can follow the same motivation,
but must account for a different behavior of the
maximum likelihood estimator and the resulting bias.
Intuitively, the penalty adjusts for an expected increase in the
maximized log-likelihood when variables are added to the model, which
is less likely under constraints. The likelihood
of violation of the constraints must be taken into account.

\citet{hughes2003} adapted the AIC criterion using the asymptotic
distribution of the Wald test statistic. In the constrained situation
this statistic is not distributed as a
chi-squared random variable anymore, but as a probability
weighted mixture of chi-squared random variables [see
\citet{chernoff1954,gourieroux1982,kodde1986} or \citet{vdv98}, Theorem~16.7].
It is of the form (partially
inequality constrained Wald statistic)
%
\begin{equation}
\label{eq4} {\sum_{h=0}^{p_{r}}{w(p_{r},h)}
\chi^{2}(k_{r}-p_{r}+h)},
\end{equation}
where $p_{r}$ is the number of inequality constraints and $w(p_{r},h)$
are weights [with $\sum_{h}w(p_{r},h)=1$], which can
be interpreted as the probabilities under the null hypothesis
that the constrained maximum likelihood estimator
$\widetilde\theta_r$ satisfies $h$ out of $p_{r}$ constraints.

\citet{hughes2003} propose to use the one-sided AIC (OSAIC), which is
an asymptotically unbiased estimator of the KL divergence in the
presence of one-sided information:
%
\begin{equation}
\label{eq5} \mathrm{OSAIC}_{r}=-\mathcal{L}_{r}(
\widetilde{\theta}_{r}) + {\sum_{h=0}^{p_{r}}{w(p_{r},h)
(k_{r}-p_{r}+h)}}.
\end{equation}
Calculating the weights is a combinatorial problem,
which aims to determine the probability that the vector $\widetilde
\theta_r$
lies in any face of dimension $h$ [\citet{kudo1963,shapiro1988,gromping2010}].
This can be computationally intensive as the number of variables, $k_r$,
increases [\citet{gromping2010}]. However, in this study the largest model
has eight free parameters (because $S\leq4$).
Therefore, the model selection procedure is still very fast (a couple
of seconds).

\subsection{Testing}
\label{methtest}
To evaluate the effect of DNA copy number on expression, we
test the hypothesis $H_{0}\dvtx \mathbf{C}\theta= 0$ against the
alternative $H_{1}\dvtx \mathbf{C}\theta\neq0,
\mathbf{C}\theta\geq0$, that is, we test that all inequality
constraints
are satisfied as equalities against the possibility
that at least one of them is strict.
From \eqref{constr} we observe that all parameters
except the intercept $\theta_{0}$ are subject to inequality constraints
and that the null hypothesis reduces the model
to the intercept.

We employ the likelihood ratio statistic $\mathrm{LR} =
2(\mathcal{L}_{1}-\mathcal{L}_{0})$, where $\mathcal{L}_{0}$ and
$\mathcal{L}_{1}$ are the maximized log-likelihood under
the null and alternative hypotheses, respectively. The test rejects the null
hypothesis for large values of
%
\begin{equation}
\label{eq6} \mathop{\operatorname{min}}_{\mathrm{C}\theta\geq0} (y-\mathbf{X}
\theta)^{T} (y-\mathbf{X}\theta) - \mathop{\operatorname{min}}_{\mathrm{C}\theta=
0}
(y-\mathbf{X}\theta)^{T} (y-\mathbf{X}\theta).
\end{equation}
This can be shown [\citet{robertson1988}] to be equivalent to
rejecting for large values of
%
\begin{equation}
\label{eq8} \overline{\chi}^{2} = (\widetilde{\theta} - \widetilde{
\theta}_{=})^{T} \Sigma^{-1}_{\mathbf{X}} (
\widetilde{\theta} - \widetilde{\theta}_{=}),
\end{equation}
where $\widetilde{\theta}$ and $\widetilde{\theta}_{=}$
are the maximum likelihood estimators
under the inequality and the equality constraints,
respectively, and $\Sigma_{\mathbf{X}}
=\sigma^2(\mathbf{X}^T\mathbf{X})^{-1}$ is
the covariance matrix of the unconstrained least squares estimator. For known
error variance $\sigma^2$ the chi-bar-squared statistic $\overline{\chi}^{2}$
may be employed with null distribution approximated
by a weighted mixture of $\chi^{2}$
distributions [\citet{chernoff1954,gourieroux1982}]. As $\sigma^2$
is typically unknown, we use instead the so-called
\emph{\mbox{E-bar}-squared statistic}
[\citet{robertson1988,shapiro1988,gromping2010,silvapulle2004}]
%
\begin{equation}
\label{eq9} \overline{E}^{2} = \frac{(\widetilde{\theta} - \widetilde{\theta}_{=})^{T}
\Omega_{\mathbf{X}}^{-1} (\widetilde{\theta} -
\widetilde{\theta}_{=})}{(\widetilde{\theta} - \widetilde{\theta}_{=})^{T}
\Omega_{\mathbf{X}}^{-1} (\widetilde{\theta} - \widetilde{\theta}_{=})+
(y-\mathbf{X}\hat{\theta})^{T} (y-\mathbf{X}\hat{\theta})}.
\end{equation}
Here $\Omega_{\mathbf{X}}=\mathbf{X}^T\mathbf{X}$.
The null distribution of this statistic is a weighted mixture of Beta
distributions of the form
%
\begin{equation}
\label{eq11} {\sum_{h=0}^{p}{w(p,h)}
\mathcal{B}\bigl(h/2,(n-p)/2\bigr)},
\end{equation}
where $p$ is the number of parameters and $\mathcal{B}(a,b)$ refers to
a beta distribution with shape parameters $a$ and $b$.
The mixing weights are the same as in \eqref{eq4} (applied to
the full model); unknown parameters are estimated by their MLEs.

Further details on these test statistics can be found in
\citet{shapiro1988,robertson1988,silvapulle2004}.

\subsection{Confidence bands}
\label{methCB}
Confidence bands (CBs) for the (spline) function
$\mathbf{x}\mapsto f_{\alpha}(\mathbf{x};\theta)$
in equation \eqref{eq1}
should take both the model selection procedure
[\citet{buckland1997}] and the constraints
into account.

Initially we implemented a bootstrap procedure [\citet{gromping2010}],
accounting for model uncertainty along the lines
of \citet{burnham2002}, who propose the
construction of so-called unconditional confidence intervals where
only the selected model is considered for each bootstrap sample.
Unfortunately, simulated coverage
probabilities were below (and sometimes far below, e.g., 0.6
instead of
0.95) the nominal level, probably due to the presence of
the inequality constraints in our model [\citet{andrews2000}]. We therefore
developed an ``exact'' alternative based on the
E-bar-squared statistic \eqref{eq9},
using semidefinite programming to achieve computational efficiency.
A simulation study reported in Section~\ref{simcb} shows that this
approach yields accurate uniform CBs.

\subsubsection{Problem formulation}
\label{methCBprob}
We start by the construction of a joint confidence region for all parameters
$\theta$ in the full model, including the intercept $\theta_{0}$,
by inverting the likelihood ratio test described previously.
Analogously to equation \eqref{eq9}, define
\[
{\overline{E}^{2}(\theta) = \frac{(\widetilde{\theta} -
\theta)^{T}
\Omega_{\mathbf{X}}^{-1} (\widetilde{\theta} - \theta
)}{(\widetilde{\theta} - \theta)^{T}
\Omega_{\mathbf{X}}^{-1} (\widetilde{\theta} - \theta)+ (y-\mathbf{X}\hat{\theta})^{T}
(y-\mathbf{X}\hat{\theta})}}.
\]
Then a $(1-\alpha)\%$ confidence region $\mathcal{R}$ for $\theta$ is
%
\begin{equation}
\label{eq13} \mathcal{R}=\bigl\{\theta\dvtx \overline{E}^{2}(\theta)
\leq\mathcal {Q}_{1-\alpha}, \mathbf{C}\theta\geq0 \bigr\},
\end{equation}
where $\mathcal{Q}_{1-\alpha}$ denotes the $(1-\alpha)$-quantile of
the beta mixture distribution in \eqref{eq11}. Here we increment the
first parameter of the Beta distributions to $(h+1)/2$, because
presently we include the intercept as a parameter, whereas
before it was free under the null hypothesis.
Interval estimation based on inversion of a likelihood
ratio statistic is known to possess good properties
[\citet{meeker1995,arnold1998,brown2003}].

Given the confidence region $\mathcal{R}$, we compute a confidence band
by determining for each $\mathbf{x}$ the minimum and maximum values
$f_{\alpha}(\mathbf{x};\theta)=\mathbf{x}^{T}\theta$.
This means determining
\[
\inf_{\theta\in\mathcal{R}} \mathbf{x}^{T}\theta \quad\mbox{and}\quad
\sup_{\theta\in\mathcal{R}} \mathbf{x}^{T}\theta.
\]
Thus, a simple linear function must be minimized (or maximized)
subject to linear and ellipsoidal inequality constraints. In the
following section, we show that this (convex) problem can be
solved efficiently by semidefinite programming.

\subsubsection{Semidefinite programming}
\label{methCBsdp}
A semidefinite program [\citet{vandenberghe1996}] is concerned with the
minimization of a linear objective function under the constraint that
a linear combination of symmetric matrices is positive semidefinite:
%
\begin{equation}
\label{opt2} \mathop{\mathrm{minimize}}_{y\in\mathbb{R}^{m}} b^{T}y
\qquad\mbox{subject to } F(y)=F_{0}+\sum_{i=1}^{m}{y_{i}F_{i}}
\succeq0.
\end{equation}
The vector $b\in\mathbb{R}^{m}$ and the symmetric $(n \times n)$
matrices $F_{0},\ldots,F_{m}$ are fixed, and the expression
$F(y)\succeq0$ means that the matrix $F(y)$ is positive semidefinite
[i.e., $z^{T}F(y)z\geq0$, $\forall z \in\mathbb{R}^{n}$]. Because a
linear matrix inequality constraint $F(y)\succeq0$ is convex, the
program can be solved efficiently using interior-point methods
[\citet{vandenberghe1996}].

We may express the optimization problem of the previous section as a
semidefinite program, based on two equivalences,
given by \citet{vandenberghe1996} and provided in
Appendix~\ref{appendsdp}. For convenience, we replace
the ellipsoidal constraint
$\overline{E}^{2}(\theta) \leq\mathcal{Q}_{1-\alpha}$
by $(M\theta- M\widetilde{\theta})^{T} (M\theta-
M\widetilde{\theta}) \leq\lambda$, where $ {\lambda=
(y-\mathbf{X}\hat{\theta})^{T}
(y-\mathbf{X}\hat{\theta})
\mathcal{Q}_{1-\alpha}/(1-\mathcal{Q}_{1-\alpha})}$ and
$\Omega_{\mathbf{X}}^{-1}=M^{T}M$. Given this, the semidefinite
program is
%
\begin{equation}
\label{opt3} \mathop{\mathrm{minimize}}_{\theta} \mathbf{x}^{T}
\theta \qquad\mbox{subject to } F(\theta)=F_{0}+\sum
_{i=1}^{p}{\theta_{i} F_{i}}
\succeq0,
\end{equation}
where
\[
F_{0}= \pmatrix{ 0 & 0 \vspace*{2pt}
\cr
0 & F^{(2)}_{0}
} ,\qquad F_{i}= \pmatrix{ F^{(1)}_{i} & 0 \vspace*{2pt}
\cr
0 & F^{(2)}_{i} } , \qquad i=1,\ldots,p,
\]
with the submatrices defined as
\[
F^{(1)}_{i}=\operatorname{diag} (c_{i}),\qquad
F^{(2)}_{0}= \pmatrix{ I & -M\widetilde{\theta} \vspace*{2pt}
\cr
(-M\widetilde{\theta})^{T} & \lambda } \quad\mbox{and}\quad
F^{(2)}_{i}= \pmatrix{ 0 & m_{i} \vspace*{2pt}
\cr
m_{i}^{T} & 0} .
\]
Here $m_{i}$ and $c_{i}$ denote the $i$th column
vector of the matrices $M$ and $\mathbf{C}$ [the matrix of linear
restrictions expressed in \eqref{opt1}], respectively.

The optimization procedure needs to be repeated twice in order to
determine the lower and upper bounds on
$\mathbf{x}^{T}\theta$. Even though this must next be repeated for
every new instance $\mathbf{x}$ to obtain a confidence band, the
overall procedure is fast. For instance, for 100 new instances
computation on a 2.66~GHz Intel quad-core took less than 12~s (without
parallel computing).

\section{Simulation}
\label{sim}
We conducted simulation experiments to: (1) determine the
accuracy of estimates as provided by PLRS (Section~\ref{simpe});
(2) examine the coverage probabilities of the method proposed in
Section~\ref{methCB}
(Section~\ref{simcb}); and (3) evaluate the performance of the PLRS
screening test in detecting associations of various functional
forms (Section~\ref{simtest}).

\subsection{Point estimation}
\label{simpe}
The simulation study examined the accuracy of the estimates
obtained by fitting piecewise splines or a simple linear model.
For simplicity, we consider a two-state model (normal and gain)
and the knot was fixed to 0.5.
Data were generated according to the following:
\begin{itemize}
\item model 1: $ y = 1 + a_2 (x-0.5)^{1}_{+}$, $a_{2}\in\{0, 0.5, 1, 2, 5\}$
\item model 2: $ y = 1 + 0.5 x + (a_2-0.5) (x-0.5)^{1}_{+}$,
$a_{2}\in\{0, 0.5, 1, 2, 5\}$.
\end{itemize}
The first state (normal) has no or little effect on expression.
The linear function is contained in both models,
and is found for $a_{2}=0$ and $a_{2}=0.5$, respectively.
We generated errors from a normal distribution $\mathcal{N}(0,\sigma^2)$
where $\sigma\in\{0.1, 0.25,0.5, 0.75, 1\}$. This resulted in 25 cases
for each of the two models (5~values of $a_2$ times 5 values of $\sigma$).
The sample size was set to 80, and the 80 values of $x$ were generated
from a uniform distribution $\mathcal{U}(0,1)$.

We were interested in comparing the precision of the estimates of the
slope $a_{2}$ when fitting a linear or a piecewise linear model (the latter
with a single knot placed at $0.5$; 4 parameters). For each of the 25 cases
we repeated the simulation experiment 1000 times and computed the
estimator of the slope for both models. Table~\ref{simEst}
reports the empirical squared bias and variance over the
1000 repetitions. For clarity only the results for $\sigma=0.25$ and
$\sigma=0.75$ are displayed. Complementary results can be found in
Section 2 of SM.

Not surprisingly, the piecewise model can capture the relationship well
in all cases: the squared bias is small, and the variance never unduly large.
On the other hand, the estimate of the slope given by the linear
model is strongly biased for larger values of the slope $a_2$.
As expected, the variance of the PLRS estimate is usually somewhat larger
than that of the linear model estimate. However, this difference is much
less prominent than for the squared bias. When the data generating
process is linear, that is, when $a_{2}=0$ in model 1 and $a_{2}=0.5$ in
model 2, the difference between the estimates from the linear and PLRS
models is smaller than in the other cases.

The study suggests that, when estimating or testing the effect of DNA copy
number on mRNA expression, there is potentially more to lose than to gain
(due to misspecification versus overspecification of the model) by applying
the linear instead of the piecewise linear spline model.

\begin{table}
\caption{Squared bias and variance (in parentheses) of the slope
estimates of the linear and piecewise spline models as a function of
the true slope $a_{2}$, noise $\sigma$ and model. In bold: setting for
which the true model is linear}\label{simEst}
\begin{tabular*}{\textwidth}{@{\extracolsep{\fill}}lccccc@{}}
\hline
&& \multicolumn{2}{c}{$\bolds{\sigma=0.25}$} & \multicolumn{2}{c@{}}{$\bolds{\sigma=0.75}$}
\\[-6pt]
&& \multicolumn{2}{c}{\hrulefill} & \multicolumn{2}{c@{}}{\hrulefill} \\
{\textbf{Model}}&{$\bolds{a_2}$}&\textbf{linear} & \textbf{piecewise} & \textbf{linear} & \textbf{piecewise} \\
\hline
{1} & \textbf{0}\phantom{0.} & \textbf{0.002 (0.004)} & \textbf
{0.007 (0.012)} & \textbf{0.015 (0.033)} & \textbf{0.047 (0.090)} \\
& 0.5 & 0.070 (0.011) & 0.002 (0.039) & 0.050 (0.060) & 0.000 (0.193)
\\
& 1\phantom{0.} & 0.282 (0.011) & 0.004 (0.045) & 0.270 (0.081) & 0.008 (0.271) \\
& 2\phantom{0.} & 1.114 (0.011) & 0.003 (0.045) & 1.124 (0.094) & 0.027 (0.339) \\
& 5\phantom{0.} & 6.962 (0.011) & 0.003 (0.042) & 6.908 (0.103) & 0.022 (0.393)
\\[3pt]
{2} & 0\phantom{0.} & 0.060 (0.008) & 0.063 (0.009) & 0.075 (0.053)
& 0.124 (0.097) \\
& \textbf{0.5} & \textbf{0.000 (0.009)} & \textbf{0.005 (0.019)} &
\textbf{0.000 (0.066)} & \textbf{0.030 (0.146)} \\
& 1\phantom{0.} & 0.058 (0.008) & 0.000 (0.036) & 0.055 (0.070) & 0.006 (0.180) \\
& 2\phantom{0.} & 0.545 (0.008) & 0.000 (0.041) & 0.521 (0.075) & 0.000 (0.289) \\
& 5\phantom{0.} & 4.782 (0.008) & 0.000 (0.046) & 4.857 (0.073) & 0.004 (0.320) \\
\hline
\end{tabular*}
\end{table}
%

\subsection{Uniform CBs}
\label{simcb}
To study the coverage probabilities of the method proposed in
Section~\ref{methCB},
we simulated data according to the model $y = 1 + (x-0.5)^{0}_{+} +
(x-0.5)^{1}_{+}$,
with $x$-values drawn from a uniform distribution
$\mathcal{U}(0,1)$. Gaussian errors of standard deviation
$\sigma\in\{0.5,1\}$, and three sample sizes $n \in\{20,40,80\}$.
For a given data set we computed the confidence band
on a grid of 10 equidistant values, for two
different significance levels $\alpha\in\{0.05, 0.1\}$,
and checked whether the 10 corresponding values of the function in the display
fall simultaneously into the estimated confidence band.
(For computational reasons the simulation was limited to 10 values; we
believe that using the continuous
range would not have altered the findings.)
Table~\ref{simCov} shows the empirical coverage probabilities
over 10,000 data sets for each situation.

The simulated coverage probabilities are close
to their corresponding nominal values. Even though the coverage
procedure is motivated by asymptotic approximations, this is true even when
the sample size is small, in agreement with
previous literature on likelihood-based interval estimation.

\begin{table}
\tablewidth=250pt
\caption{Simulated coverage probability for different sample sizes,
noise levels and significance levels}\label{simCov}
\begin{tabular*}{250pt}{@{\extracolsep{\fill}}lcccc@{}}
\hline
& \multicolumn{2}{c}{$\bolds{\sigma=0.5}$} & \multicolumn{2}{c@{}}{$\bolds{\sigma=1}$}
\\[-6pt]
& \multicolumn{2}{c}{\hrulefill} & \multicolumn{2}{c@{}}{\hrulefill} \\
& \multicolumn{1}{c}{$\bolds{\alpha=0.05}$} & \multicolumn{1}{c}{$\bolds{\alpha=0.1}$} &
\multicolumn{1}{c}{$\bolds{\alpha=0.05}$} & \multicolumn{1}{c@{}}{$\bolds{\alpha=0.1}$} \\
\hline
$n=20$ & 0.953 & 0.898 & 0.968 & 0.922 \\
$n=40$ & 0.952 & 0.883 & 0.967 & 0.926 \\
$n=80$ & 0.939 & 0.863 & 0.960 & 0.915 \\
\hline
\end{tabular*}
\end{table}

\subsection{PLRS screening test}
\label{simtest}
We evaluated the performance of the PLRS
testing procedure in detecting associations of various functional
shapes.
PLRS was compared to the LM test (see Section~\ref{apptesting}),
Spearman's correlation test and the test proposed
by \citet{vanWieringen2009}.
SM Figures 2 to 11 show partial ROC curves (sensitivity versus
type I error $\alpha$, where $\alpha\leq0.2$) and partial AUC.
Details are provided in SM Section 4. Here, we
summarize the results.

The PLRS test yielded good performance in detecting various
types of associations. It achieved the highest AUC in
68 out of the 90 simulation cases (against 23 for LM).
When the true effect is linear, PLRS performed reasonably well.
In other cases, it always produced a high, if not the highest, AUC.
In particular, PLRS presented a clear advantage over others in detecting
partial effects on gene expression, that is, when only one abnormal state
(among others) affects expression.
In all, results suggest that PLRS accommodates well both continuous
and discrete genomic information and, unlike others, is able to detect
various types of association.

\section{Application}
\label{app}
The proposed framework was applied to two data sets.
The first data set [\citet{carvalho2009}; available at \href{http://www.ncbi.nlm.nih.gov/geo/}{ncbi.nlm.nih.gov/geo};
accession number GSE8067] consists of copy number and gene expression
values for 57 samples of colorectal cancer tissue. These were generated with
BAC/PAC and Human Release 2.0 oligonucleotide arrays, respectively.
Normalization is as in \citet{carvalho2009}. aCGH data were segmented with
the CBS algorithm of \citet{olshen2004} and discretized with CGHcall
[\citet{vandeWiel2007}]. Matching of mRNA and aCGH features was
based on minimizing the distance between the midpoints of the genomic
locations of the array elements. The final data set comprises 25,869 matched
features. The second data set [\citet{neve2006}; available from Bioconductor]
consists of copy number and expression data for 50 samples (cell
lines) of breast cancer, profiled with OncoBAC and Affymetrix HG-U133A
arrays. Preprocessing of mRNA expression is described in \citet{neve2006}.
aCGH data were segmented and called as above. The resulting data set
contains 19,224 matched features. For the colorectal and breast cancer
data sets,
knots of the PLRS model were estimated using methods I and II, respectively.

We first present some global results on model selection, and next
consider testing the association between DNA and mRNA. Finally, some
relevant relationships are illustrated.

\subsection{Model selection with the OSAIC procedure}
\label{appmodSel}
Table~\ref{appModSel} reports the number of genes for which our procedure
(column OSAIC) selects a certain type of model, for both data sets.
Clearly, both the piecewise linear model and the piecewise level model
are selected a large number of times.
Different procedures such as AIC and BIC,
$\mathrm{BIC}_{r}=-2\cdot\mathcal{L}_{r}(\widetilde{\theta}_{r})+
\log(n)\cdot k_{r}$, which put bigger penalities on larger models
(too large given the constraints), still often prefer piecewise
splines. This gives strong evidence on the inadequacy of both the
simple linear and piecewise constant models for many genes.
In Section 1 of SM, an overlap comparison of the three
procedures shows differences induced by the different penalty functions.

\begin{table}
\caption{The number of times a model is selected
by type of model, by three model selection procedures,
for the two data sets}\label{appModSel}
\begin{tabular*}{\textwidth}{@{\extracolsep{\fill}}lcccccc@{}}
\hline
& \multicolumn{3}{c}{\textbf{\citet{carvalho2009}}} & \multicolumn{3}{c@{}}{\textbf{\citet{neve2006}}}
\\[-6pt]
& \multicolumn{3}{c}{\hrulefill} & \multicolumn{3}{c@{}}{\hrulefill} \\
\textbf{Type of model} & \textbf{OSAIC} & \textbf{AIC} & \textbf{BIC} & \textbf{OSAIC} & \textbf{AIC} &
\textbf{BIC}\\ \hline
Intercept & \textbf{14,720} & 18,083 & 21,700 & \textbf{5081} & 6968 &
9379\\
Simple linear & \phantom{0.}\textbf{4916} & \phantom{0.}3674 & \phantom{0.}2043 & \textbf{5262} & 6689 &
6345\\
Piecewise level & \phantom{0.}\textbf{2667} & \phantom{0.}1977 & \phantom{00.}992 & \textbf{2761} & 2477 &
1608\\
Piecewise linear & \phantom{0.}\textbf{3566} & \phantom{0.}2135 & \phantom{0.}1134 & \textbf{6120} & 3090
& 1892\\
\hline
\end{tabular*}
\end{table}

\subsection{Testing the effect of DNA on mRNA}
\label{apptesting}
The hypothesis that DNA copy number has no effect on mRNA expression
corresponds to model \eqref{eq1} with only the intercept parameter
$\theta_0$ nonzero. We tested this as the null model both versus the~full
model \eqref{eq1} (test ``PLRS'') and versus the linear submodel
(test ``LM''), with the purpose to compare these two screening models in
their effectiveness to detect an association. A third possibility
would be to test the null model versus the model selected by the OSAIC
procedure. However, because this would naively suggest that the form
of the relationship is known a priori, we did not pursue this option.
For the PLRS test a minimum number of five observations (the default
being three) per state was imposed.

\begin{table}
\caption{Number of associations with an estimated false discovery rate
below 0.1 for different model comparisons}\label{appFDR}
\begin{tabular*}{\textwidth}{@{\extracolsep{\fill}}lccc@{}}
\hline
$\bolds{H_{0}}$ & $\bolds{H_{a}}$ & \textbf{\citet{carvalho2009}} & \multicolumn{1}{c@{}}{\textbf{\citet{neve2006}}} \\
\hline
Intercept & linear & 1726 & 9783 \\
Intercept & full & 1554 & 9105 \\
\hline
\end{tabular*}
\end{table}

\begin{figure}[b]

\includegraphics{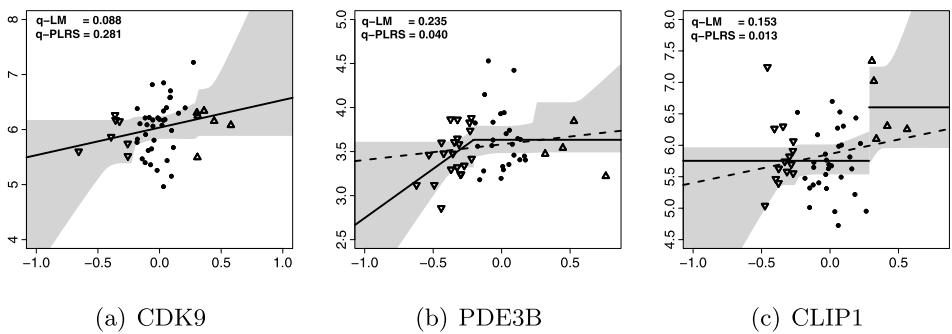}

\caption{Association between DNA and mRNA for different genes
in the breast cancer data set [\citet{neve2006}]. Segmented
copy number is on the x-axis, while gene expression is on the
y-axis. Symbols indicate the different states, namely,
loss ($\bigtriangledown$), normal ($\bigcirc$) and gain
($\bigtriangleup$). Grey surfaces correspond to 95\%
uniform CBs. The top left values correspond to $q$-values
of tests LM and PLRS, respectively. The dashed line gives
the fitted LM model; the ``continuous'' spline is the
fitted PLRS model.}
\label{smalleffects}
\end{figure}

Table~\ref{appFDR} gives the number of associations with a $q$-value below
0.1 [based on the \citet{benjamini1995} FDR]. The LM test is seen to
detect slightly more associations as being significant than the PLRS
test. This may be a consequence of the fact that the linear model
involves fewer parameters. However, closer inspection shows that the
sets of detected genes are not nested, and the PLRS test is able to
detect biologically meaningful genes that are not detected by the LM
test. To
illustrate, three DNA-mRNA relationships are plotted in
Figure~\ref{smalleffects}. The first corresponds to an association
detected as
significant with the LM test, but not with the PLRS test. Reciprocally, the
last two associations (genes PDE3B and CLIP1) are detected with the PLRS
test but not with the LM test. The figure shows that the PLRS
test is able to detect relationships for which an effect is present
for only a few samples (but at least five). Identifying
the last two genes may be more important than the first,
as they are more interesting potential targets for studying
individual effects.

The first gene in Figure~\ref{smalleffects} also illustrates that the testing
procedures may differ considerably in $q$-values, even though the
estimated regression function found by the two models is the same.
This is partly explained by the difference in complexity between the
alternative models. However, we note that $q$-values for a single gene
are not directly comparable, since they also depend on $p$-values of
other genes. In Appendix~\ref{appendtest}, we provide, for selected
genes, $p$- and $q$-values for the different types of tests.

%

\subsection{Results for selected genes}
\label{appassoc}
In this section we show the estimated relationships for selected
genes. The selection is based on the Cancer Gene Census list
(available at \href{http://www.sanger.ac.uk/genetics/CGP/Census/}{www.sanger.ac.uk/genetics/CGP/Census/}) and on our
observation that some associations are atypical. Also, we show results
for genes C20orf24, TCFL5 and TH1L, which were reported in
\citet{carvalho2009} as important for colorectal cancer progression.

\begin{figure}

\includegraphics{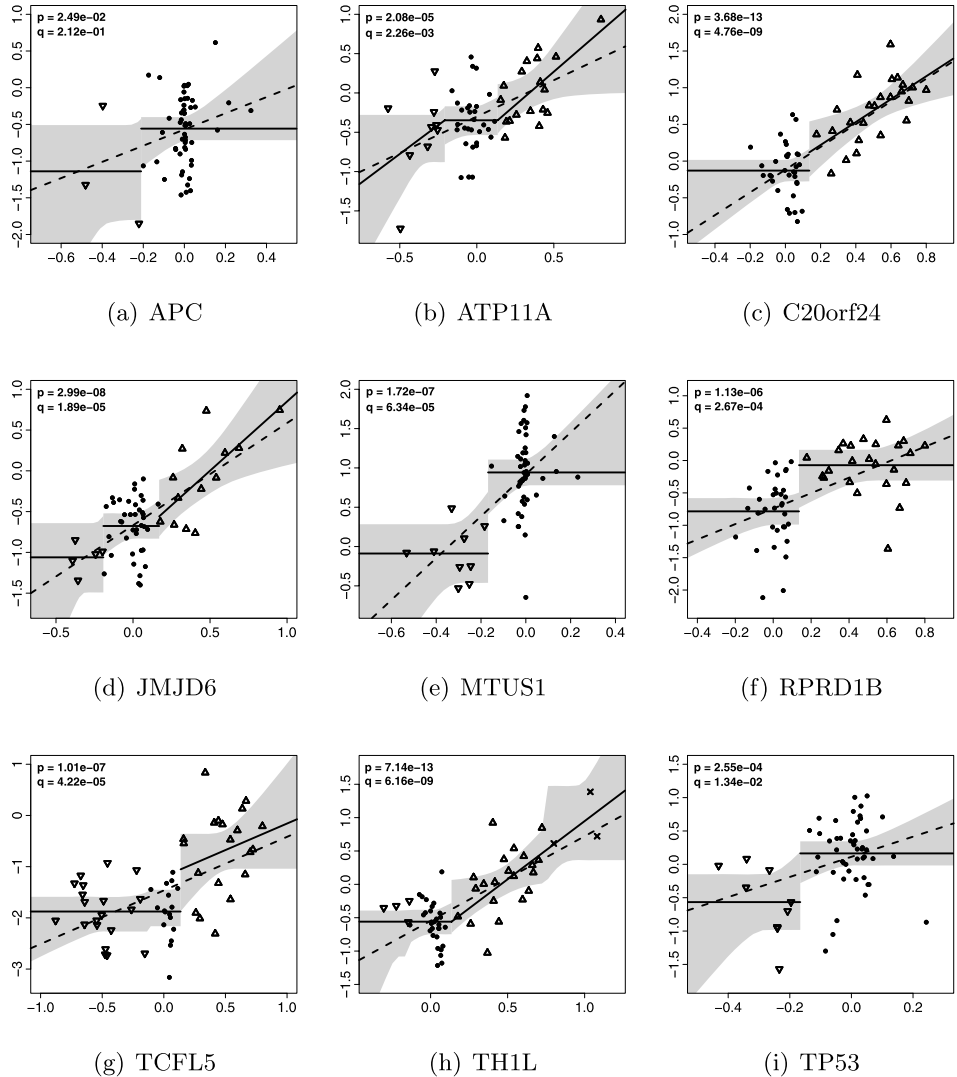}

\caption{Association between DNA and mRNA for different genes in the
colorectal cancer data set. Segmented copy number is on the x-axis, while
gene expression is on the y-axis. States are indicated by different symbols:
loss ($\bigtriangledown$), normal ($\bigcirc$), gain
($\bigtriangleup$) and amplification ($\times$). Grey surfaces
correspond to 95\% uniform CBs. In all cases the piecewise linear
model is preferred to the simple linear one (dashed line). The top
left values correspond to the $p$- and $q$-values of the PLRS test.}
\label{carvalho}
\end{figure}

\begin{figure}

\includegraphics{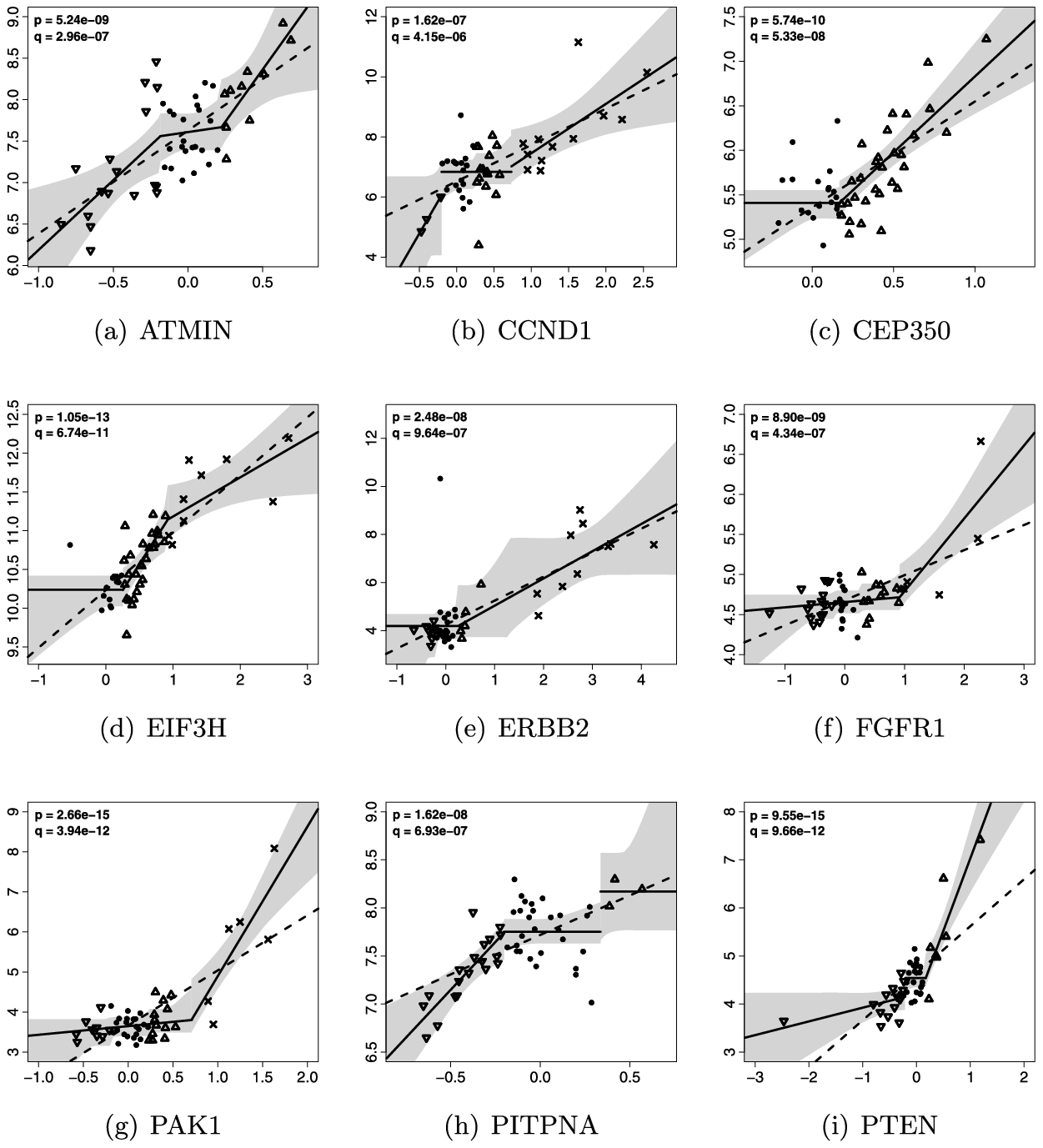}

\caption{Association between DNA and mRNA for different genes in the
breast cancer data set. Segmented copy number is on the x-axis, while
gene expression is on the y-axis. States are indicated
by different symbols: loss ($\bigtriangledown$), normal ($\bigcirc$), gain
($\bigtriangleup$) and amplification ($\times$). Grey surfaces
correspond to 95\% uniform CBs. In all cases the piecewise linear
model is preferred to the simple linear one (dashed line). The top
left values correspond to the $p$- and $q$-values of the PLRS test.}
\label{neve}
\end{figure}

Figures~\ref{carvalho} and~\ref{neve} show nine DNA-mRNA associations for
each of the two data sets. Each plot displays the fit of the linear
model and of the PLRS model chosen by the OSAIC
criterion. Uniform 95\% confidence bands (that account for model
selection uncertainty) are also plotted. (Some curious shapes
result from the fact that pointwise variation
bursts near the boundaries and around knots.)

%

Both figures show a diverse set of forms of associations. Fitted models
with jumps reveal that discrete copy number states can, by themselves, explain
variation in expression. This is even more true when a piecewise level
relationship is identified (as for genes APC and MTUS1 in
Figure~\ref{carvalho}). More generally, piecewise linear models
capture effects that differ for losses, gains and/or
amplifications. Statistically speaking, this has the advantage of
giving more accurate estimates of slope(s), as is clearly observed for
genes ATMIN, PITPNA and PTEN in Figure~\ref{neve}. Having a better estimator,
we may expect a better test. From a biological point of view, the
ability to distinguish effects between states may help the detection of
onco and tumor-suppressor genes. Moreover, genes for which these effects
concern only a few samples may also be interesting to biologists for
studying individual effects.

The simple linear model is observed to be a tight template for
modeling. As a matter of fact, it is potentially misleading when the
relationship really depends on the underlying copy number state. This
happens to be the case for known cancer genes (see FGFR1, PAK1 and
PTEN in Figure~\ref{neve}). As a result, when testing the effect of
DNA on mRNA with the LM and PLRS tests (see
Section~\ref{apptesting}), one may obtain a considerable difference
between the $p$-values, and hence $q$-values (see
Appendix~\ref{appendtest}). For this reason the proposed framework
may improve the detection of (highly) significant associations and
their ranking.

Finally, we dwell on the notion of effect in itself.
The notion of ``association'' is broad, and can be expressed both
by an intercept and a slope. This can
imply a clear difference\vadjust{\goodbreak} in interpretation with respect to the linear
model. Consider the simple example of gene MTUS1 in Figure~\ref{carvalho},
where a piecewise level model is preferable. Here intuition clearly
tells us that one is more interested in assessing the difference in
expression level between samples presenting loss and normal
aberrations than an overall trend. Therefore, a linear model
may focus on the wrong quantity of interest, whereas the PLRS
procedure may yield meaningful interpretation.

We concentrated on comparing our results with those of the linear
model. However, it is clear from Figures~\ref{carvalho} and~\ref{neve}
that also the other alternative, the piecewise level model (which
allows only horizontal lines per state), is often not adequate (see
TH1L and PITPNA).

\section{Conclusion}
We proposed a statistical framework for the integrative analysis of
DNA copy number and mRNA expression, which incorporates segmented and
called aCGH data. By using discrete aCGH data we improved model
flexibility and interpretability. The form of the relationship is
allowed to vary per gene. Model interpretation is ameliorated with
biologically motivated constraints on the parameters. This complicates
the statistical procedures for identifying and inferring the
relationship between the markers, but we provided methods for model
selection, interval estimation and testing the strength of the
association. We applied the methodology to two real data sets. Many
(reported) genes exhibited interesting behavior.

A novelty of this work is the combined use of segmented and called
aCGH data. Which of the two data types is more suitable is a matter of
debate in the aCGH community, and may depend on the type of
downstream analysis [\citet{vanWieringen2007}]. Our method provides a
compromise that uses both characteristics of the data.

The form of association between copy number and expression in breast
cancer is also explored in the recent paper \citet{solvang2011} (which
we received after completion of this paper). This interesting paper
distinguishes (only) between linear and quadratic types of effect and
uses (only) two types of aberrations, without distinguishing gains
from amplifications. The interpretation of the coefficients in our
model seems to be simpler.

The proposed methodology is also applicable to the joint analysis of
copy number and microRNA expression. This class of noncoding RNA was
shown to play an important role in tumor development. Our method may
be particularly suitable for these data, because microRNA transcripts are
often expressed in part of the samples only.

Next generation sequencing data will impose new challenges, which will
be taken up in future work. This type of data provides higher
resolution than microarrays, while reducing biases, in particular, at
the lower end of the spectrum. Because expression levels are measured
as counts rather than intensities, the distribution of the response
variable cannot be assumed to be Gaussian and, hence, a different
noise model is needed.

In short, we provide methodology for statistical inference and model
selection in the framework of constrained PLRS, and show that this is
able to reveal interesting DNA-mRNA relationships for cancer
genes. The method is implemented in R and available as a package
from \href{http://www.few.vu.nl/\textasciitilde mavdwiel/software.html}{www.few.vu.nl/\textasciitilde mavdwiel/}
\href{http://www.few.vu.nl/\textasciitilde mavdwiel/software.html}{software.html}.

\begin{appendix}

\section{Testing}\label{appendtest}

See Table \ref{appendTest}.

\begin{table}
\def\arraystretch{1.1}
\caption{$p$ and $q$-values of the test when under the alternative
hypothesis H$_{a}$ the linear, OSAIC-selected and the full models are
successively considered. The top and bottom parts correspond,
respectively,
to the selected genes from the colorectal and breast data}\label{appendTest}
\begin{tabular*}{\textwidth}{@{\extracolsep{\fill}}lcccccc@{}}
\hline
& \multicolumn{2}{c}{\textbf{Linear}} & \multicolumn{2}{c}{\textbf{OSAIC}\phantom{00.}} & \multicolumn
{2}{c@{}}{\textbf{Full}} \\[-6pt]
& \multicolumn{2}{c}{\hrulefill} & \multicolumn{2}{c}{\hrulefill\hspace*{7.5pt}} & \multicolumn
{2}{c@{}}{\hrulefill} \\
& $\bolds{p}$ & $\bolds{q}$ & $\bolds{p}\phantom{00.}$ & $\bolds{q}\phantom{00.}$ & $\bolds{p}$ & $\bolds{q}$ \\
\hline
APC & 2.49e-02 & 2.04e-01 & 2.26e-02 & 6.38e-02 & 2.49e-02 & 2.12e-01
\\
ATP11A & 7.34e-06 & 9.79e-04 & 5.88e-06 & 2.98e-04 & 2.08e-05 &
2.26e-03 \\
C20orf24 & 1.71e-12 & 2.21e-08 & 3.06e-13 & 1.14e-09 & 3.68e-13 &
4.76e-09 \\
JMJD6 & 5.44e-09 & 4.85e-06 & 1.78e-08 & 4.31e-06 & 2.99e-08 &
1.89e-05 \\
MTUS1 & 6.83e-07 & 1.77e-04 & 6.38e-08 & 1.06e-05 & 1.72e-07 &
6.34e-05 \\
RPRD1B & 3.18e-06 & 5.45e-04 & 5.17e-07 & 5.02e-05 & 1.13e-06 &
2.67e-04 \\
TCFL5 & 6.49e-06 & 8.88e-04 & 1.75e-08 & 4.31e-06 & 1.01e-07 &
4.22e-05 \\
TH1L & 1.06e-10 & 3.25e-07 & 2.72e-13 & 1.14e-09 & 7.14e-13 & 6.16e-09
\\
TP53 & 6.54e-03 & 9.87e-02 & 9.42e-05 & 2.25e-03 & 2.55e-04 & 1.34e-02
\\ [3pt]
ATMIN & 1.12e-09 & 6.45e-08 & 1.13e-09 & 4.56e-08 & 5.24e-09 &
2.96e-07 \\
CCND1 & 1.91e-08 & 5.71e-07 & 3.56e-08 & 6.88e-07 & 1.62e-07 &
4.15e-06 \\
CEP350 & 8.55e-08 & 1.93e-06 & 3.07e-10 & 1.69e-08 & 5.74e-10 &
5.33e-08 \\
EIF3H & 1.70e-12 & 4.88e-10 & 8.22e-15 & 3.75e-12 & 1.05e-13 &
6.74e-11 \\
ERBB2 & 4.46e-10 & 3.18e-08 & 4.34e-10 & 2.15e-08 & 2.48e-08 &
9.64e-07 \\
FGFR1 & 1.62e-06 & 2.03e-05 & 3.99e-10 & 2.02e-08 & 8.90e-09 &
4.34e-07 \\
PAK1 & 1.15e-10 & 1.21e-08 & $<$2.2e-16\phantom{00.} & $<$2.2e-16\phantom{00.} & 2.66e-15 &
3.94e-12 \\
PITPNA & 1.85e-06 & 2.25e-05 & 8.40e-10 & 3.66e-08 & 1.62e-08 &
6.93e-07 \\
PTEN & 7.27e-09 & 2.64e-07 & 9.10e-15 & 4.02e-12 & 9.55e-15 & 9.66e-12
\\
\hline
\end{tabular*}
\end{table}

\section{Semidefinite programming}
\label{appendsdp}
Here, we provide the two equivalence relationships from
\citet{vandenberghe1996} that are necessary to express the
semidefinite program. We recall that a linear matrix inequality (LMI)
type of constraint includes, among others, linear and convex quadratic
inequalities. These are the two types of constraints we are interested
in. To express them as two LMIs, we make use of the following
equivalences.

A linear inequality constraint $Ax+b \geq0$, where
$A=  [ a_{1}\cdots a_{k}  ]$ and $x \in\mathbb{R}^{n}$, is
equivalent to the following LMI:
\[
F(x)=F_{0}+\sum_{i=1}^{k}{x_{i}F_{i}}
\succeq0,
\]
where $F_{0} = \operatorname{diag} ( b )$, $F_{i}=\operatorname{diag} ( a_{i} )$,
$i=1,\ldots,k$.
$\operatorname{diag}(v)$ represents the diagonal matrix with the vector $v$ on
its diagonal.

A convex quadratic constraint $(Ax+b)^{T}(Ax+b) - c^{T}x - d \leq0$,
where $A= [a_{1}\cdots a_{k}]$ and $x \in\mathbb{R}^{n}$, is equivalent
to the following LMI:
\[
F(x)=F_{0}+\sum_{i=1}^{k}{x_{i}F_{i}}
\succeq0,
\]
where
\[
F_{0}= \pmatrix{ I & b \vspace*{2pt}
\cr
b^{T} & d } ,\qquad
F_{i}= \pmatrix{ 0 & a_{i} \vspace*{2pt}
\cr
a_{i}^{T} & c_{i} } ,\qquad i=1,\ldots,k.
\]
Multiple LMIs can be expressed as a single one using block diagonal
matrices [\citet{vanAntwerp2000}].
\end{appendix}

\section*{Acknowledgments}
We wish to thank Thang V. Pham for helpful
discussions on optimization.

\begin{supplement}
\stitle{Complementary results and simulations}\label{suppA}
\slink[doi]{10.1214/12-AOAS605SUPP} 
\sdatatype{.pdf}
\sfilename{aoas605\_supp.pdf}
\sdescription{We present a simulation study which
compares the performance of the PLRS testing
procedure in detecting associations of various
functional shapes with that of other procedures.
Additionally, we provide an overlap comparison of model selection
procedures, complementary results for the simulation on point
estimation and a description of the simulation on the precision
of knots.}
\end{supplement}




\printaddresses

\end{document}